# The Quantum-Relativistic Dynamics of the Universe

Vladimír Skalský

Faculty of Materials Science and Technology of the Slovak University of Technology, 917 24 Trnava, Slovakia; e-mail: vladimir.skalsky@stuba.sk

**Abstract.** From the analysis of the quantum and relativistic properties of the particles it results the unified quantum-relativistic dynamics of the physical reality (Universe).

> "…may be the world is what we've always known: the Standard Model and general relativity."
>
> Steven Weinberg [17]

## 1 Introduction

At the beginning of the last century two complementary physical theories, which describe the physical reality, originated: the *Einstein general relativity* (GR), which describes the *macro-world* and the *quantum mechanics* (QM), which describes the *micro-world*.

The *physical reality* (*Universe*) is *quantum-relativistic*, i.e. the relativistic macro-world in a final result is composed by the quantum objects (particles), and vice versa, the particles can exist only in the relativistic macro-world [13, 15, 16].

The macro-world is described by the GR, and its two special partial solutions: the *Einstein theory of special relativity* (SR), and the *Newton theory of general gravitation*.

The micro-world is described by the QM, the *quantum field theory* (QFT), which is a combination of QM and SR, and three quantum theories, which describe three of the four *fundamental physical interactions* (*forces*): the *quantum electrodynamics* (QED), *electroweak theory*, and *quantum chromodynamics* (QCD), which are on the basis of the *calibration symmetry* connected into a *unified complex metatheory*, known as the *grand unified theories* (GUTs).

The QFT is the foundation of the modern *theory of particle physics*. In 1960-2000 the *Standard Model of particle physics* (in short *Standard Model*), was elaborated, the component of which is the *Standard Model of Interactions* (*forces mediated by gauge bosons*).

We know four basic *physical interactions* (*forces*): *electromagnetic*, *weak nuclear*, *strong nuclear*, and *gravity*.

The electromagnetic, weak nuclear and strong nuclear forces are mediated by the exchange particles (gauge bosons).

The *electromagnetic forces* are mediated by electrically neutral *photons* with a zero own (rest) mass.

The *weak nuclear forces* are mediated by three kinds of gauge bosons: the electrically neutral *bosons* $Z^0$, *bosons* $W^+$ with a positive charge, and *bosons* $W^-$ with a negative charge.

The *strong nuclear forces* are mediated by eight kinds of *gluons*, which transfer three kinds of colour charges.

The *gravitational* (more precisely *inertial-gravitational*), *forces* are the resultant of the quantum fluctuation of all real and virtual particles in the local space-time (field) [16].

## 2  An analysis of the quantum and relativistic properties of the particles

The correlation of the quantum and relativistic properties of particles can be understood using the analysis of the complementary quantum and relativistic properties of particles. Using this analysis we will study how the particles behave according to GR and according to QM [16].

At the beginning we will analyze how a particle behaves according to SR (which is a special partial solution of GR), and how it behaves according to QM (i.e. *de facto* we will analyze how the particle behaves according to QFT, which is a combination of QM and SR).

If the particle is at rest, then—from a viewpoint of SR—it constantly remains in the same place. From a viewpoint of QM the particle as a result of the *Heisenberg uncertainty principle* (*relations*) [10] permanently *fluctuates*. However, if we observe the particle for a sufficiently long time, we find that the particle fluctuates around a certain resultant place, which is identical with the place, at which we observe it from the viewpoint of SR.

According to SR, for the particle at rest, the *law of inertia* is valid.

The presentation of the particle at rest according to SR is different from its presentation according to QM only in that we in it abstract from its quantum fluctuations. From this fact and the fact that the relativistic macro-world is—in the final result—composed by the quantum objects unambiguously results: *The inertial properties of particles are the result of it quantum fluctuations in the local space-time* (*field*) [16].

Now let us look what will happen with the same particle at rest, when we move with respect to it by a uniform and rectilinear motion.

Because the motion is relative we will see how the particle moves with respect to us by the uniform and rectilinear motion.

From the viewpoint of SR the matter object, which moves at the uniform and rectilinear motion has special-relativistic properties [2].

Because the motion is relative, also all special-relativistic properties of the moving matter object are relative.

*For example*: The observer in the inertial system A observes the special-relativistic effects in the inertial system B, which moves with respect to him a certain constant velocity. The observer in the inertial system B—due to the relativity of motion—observes them in the inertial system A.

*Because the special-relativistic properties of the matter objects mowing towards each to other at a constant velocity are reciprocal, in principle they can not be explained by the interaction mediated by any exchange particles* [16].

In the inertial systems we cannot feel the inertia of our body. We feel it only when the system changes velocity. For example: during accelerating or slowing down of movement speeds of various vehicles.

In the gravitational field the matter objects move an accelerated motion.

Albert Einstein in 1907 noticed that if somebody falls down in a free fall in a gravitational field, he feels neither inertia nor acceleration of his body.

Einstein, on the bases of this notice, postulated the *principle of equivalence of gravity and inertia* (in short the *equivalence principle*), according to which the gravitational and inertial forces have the same physical nature and the same physical laws are valid for them.

Under an assumption that the inertial properties of particles are the result of quantum fluctuation in the local space-time (field), from the principle of equivalence of gravity and inertia it results unambiguously: *The inertial and gravitational forces* (*interactions*) *are caused by the same source*: *the quantum fluctuations of particles in the local space-time* (*field*) [16].

Between the electromagnetic, weak nuclear and strong nuclear interactions (mediated by exchange particles), and inertial-gravitational interactions (which are results of the quantum fluctuation of all particles), is a principal difference. The effects of the interactions mediated by exchange particles are *linear*. The effects of inertial-gravitational interactions, which are the results of the quantum fluctuation of all real and virtual particles in the local space-time (field), are *nonlinear*. Therefore: *Inertial-gravitational interactions in principle can not be mediated by exchange particles* [16].

The principle of equivalence enabled Einstein to generalize the *special principle of relativity* (according to which in all inertial systems the same laws of physics are valid), on the *general principle of relativity* (according to which in all inertial and non-inertial systems the same physical laws are valid).

In 1907-1915 Einstein using the general principle of relativity and equivalence principle elaborated GR.

In 1915 Einstein constructed the *gravitational field equations*, shortly *field equations* [3]:

*Einstein field equations*

$$G_{im} = -\kappa \left( T_{im} - \frac{1}{2} g_{im} T \right), \qquad (1)$$

where $G_{im}$ is the Einstein or conservative tensor, $\kappa$ Einstein gravitational constant [$\kappa = (8\pi G)/c^4$], $T_{im}$ energy-momentum tensor, $g_{im}$ metric or fundamental tensor, and $T$ scalar or trace of energy-momentum tensor ($T \equiv T_i^i$).

At present time the Einstein field equations are prevailingly presented in this version [6]:

$$R_{ik} - \frac{1}{2} g_{ik} R = -\kappa T_{ik}, \qquad (1a)$$

where $R_{ik}$ is the Riemann or curvature tensor, and $R$ scalar curvature.

In 1917 Einstein tried to apply the field equations (1) to the whole Universe.

Einstein assumed that the observed *relativistic Universe* is *spherical*, *homogeneous*, *isotropic* and *static*. The field equations (1), applied to the whole hypothetical spherical homogeneous and isotropic relativistic Universe, however, do not give a static solution. Therefore, Einstein tried to generalize (modify) them so as to give a static solution.

According to Einstein, there is only one simple possibility of modification of the field equations (1) to the equations that give the static solution and which do not violate the principles of GR. This possibility is the addition of a term of the type $kg_{im}$, where $k$ is a constant. Therefore, Einstein supplemented the field equations (1) by a *hypothetical cosmological supplement* (*member*) and thus modified them to the *theoretically* (*hypothetically*) *most general possible version of the field equations* [5]:

*Einstein modified field equations*

$$G_{\mu\nu} - \lambda g_{\mu\nu} = -\kappa\left(T_{\mu\nu} - \frac{1}{2}g_{\mu\nu}T\right), \tag{2}$$

where $\lambda$ is the Einstein cosmological constant.

The Einstein modified field equations are also presented in this form [7]:

$$\left(R_{ik} - \frac{1}{2}g_{ik}R\right) + \lambda g_{ik} = -\kappa T_{ik}. \tag{2a}$$

The cosmological constant $\lambda$ in the Einstein modified field equations (2), or (2a), can obtain values: $\lambda > 0$, $\lambda < 0$, or $\lambda = 0$. It means that under an assumption that observed homogeneous and isotropic Universe is relativistic, the Einstein modified field equations (2), or (2a), formally (mathematically) are guaranteed correct, because formally they are valid with all mathematically possible values of the cosmological constant $\lambda$, including zero.

The Einstein modified field equations (2), or (2a), can be physically valid only with one from the infinite number of mathematically possible values of the Einstein cosmological constant $\lambda$. Its value can be determined:

a) *empirically* (using astronomical observations),

b) *theoretically*.

Einstein was convinced that the *total* (*resultant*) *gravitational interaction of matter objects* in the Universe is exactly compensated by the *hypothetical cosmic repulsive force*, which is determined by the value of cosmological constant [5]:

$$\lambda = \frac{\kappa\rho}{2} = \frac{1}{r^2}, \tag{3}$$

where $\rho$ is the average mass density,

$$r = \frac{1}{\sqrt{\lambda}} \tag{4}$$

is the radius of spherical space, which has the volume

$$V = 2\pi^2 r^3. \tag{5}$$

The total mass in the Einstein model of the hypothetical spherical static homogeneous and isotropic relativistic Universe $M$ is finite, and is determined by the relations [5]:

$$M = 2\pi^2 r^3 \rho = 4\pi^2 \frac{r}{\kappa} = \frac{\sqrt{32}\pi^2}{\sqrt{\kappa^3 \rho}}. \tag{6}$$

In 1929 Edwin P. Hubble discovered the *expansion of the Universe* [11].

Hubble, on the basis of astronomical observations of the nebulae (galaxies) found "...a roughly linear relation between velocities and distances among nebulae..." [11, p. 173].

At present time this relation is known as

*Hubble law*

$$v = HD, \tag{7}$$

where $v$ is the velocity of galaxy, $D$ is its distance, and $H$ Hubble "constant" (coefficient, parameter).

Hubble's discovery of the expansion of the Universe definitely finished the static conception of the Universe and the Einstein model of the spherical static Universe was degraded to the level theoretical possibility.

Thereby, however, the destiny of Einstein's model of the spherical static Universe did not finish. In 1930 Arthur S. Eddington published the article *On the Instability of Einstein's Spherical World*, in which he proved that the Einstein modified field equations (2), applied to the whole homogeneous and isotropic relativistic Universe, do not give a static, but only a *quasi-static* solution. Therefore, the Einstein model of spherical static homogeneous and isotropic relativistic Universe is extremely unstable and any arbitrarily small fluctuation, would convert it into a *dynamic one* [1].

Einstein—under an impression of these facts—in 1931 resigned the model of spherical static Universe [7]. Later, the introduction of the cosmological constant into the relativistic cosmology he designated as the biggest blunder of his life.

From the Hubble discovery of the Universe expansion [11] it results a number of mutually bounded serious consequences, which allows to unambiguously determine the physical and model properties of our observed *expansive homogeneous and isotropic relativistic* (more precisely *quantum-relativistic*) *Universe* [13, 14, 15, 16].

*The observed expansive homogeneous and isotropic relativistic Universe*—in which the Hubble law (7) is valid—*is finite and* (throughout its whole expansive evolution), *expands at the maximum velocity of signal propagation c*.

*The expansive homogeneous and isotropic relativistic Universe*—which expands at the velocity $c$—*has the total energy $E_{tot} = 0$, in bigger distances* [than are the biggest *hierarchical gravitationally bound rotating systems* (HGRSs)], *it has special-relativistic properties, and the generally relevant laws of conservation energy, momentum, and momentum of momentum are valid in it* [13].

In 1916 Albert Einstein in the article *Die Grundlage der allgemeinen Relativitätstheorie* about two generally relevant laws of conservation wrote: "…laws of conservation of momentum and energy do not apply in the strict sense for matter alone, but they apply only in a case when the $g^{\mu\nu}$ are constant, i.e. when the field intensities of gravitation vanish." [4, p. 810].

Vladimir Fock in the book *The Theory of Space Time and Gravitation* argues: "…the absence of a gravitational field implies the absence of deviation of space-time geometry from the Euclidean and hence the vanishing of the curvature tensor. Hence, if $T^{\mu\nu} = 0$ one must also have $R^{\mu\nu} = 0$ and also $R = 0$, which is only true if the left-hand sides of the equations relating $G^{\mu\nu}$ and $T^{\mu\nu}$ do not contain a term $\lambda g^{\mu\nu}$, i.e. only if $\lambda = 0$." [9, p. 173].

These facts make it possible to unambiguously determine the exact solution of the Einstein modified field equations (2), or (2a), applied to the whole expansive homogeneous and isotropic relativistic Universe in which the generally relevant laws of conservation energy, momentum and momentum of momentum are valid [12]:

$$G_{\mu\nu} - \lambda g_{\mu\nu} = -\kappa\left(T_{\mu\nu} - \frac{1}{2}g_{\mu\nu}T\right) = 0, \tag{8}$$

where $G_{\mu\nu} = 0$, $\lambda = 0$, $T_{\mu\nu} = 0$, and $T = 0$,
or:

$$\left(R_{ik} - \frac{1}{2}g_{ik}R\right) + \lambda g_{ik} = -\kappa T_{ik} = 0, \tag{8a}$$

where $R_{ik} = 0$, $R = 0$, $\lambda = 0$, and $T_{ik} = 0$.

In 1938 Albert Einstein, Banesh Hoffmann and Leopold Infeld proved that the Einstein field equations describe not only the gravity, but also the motion of matter objects in the four-dimensional space-time (field) [8]. So they overcome the separation of the field equations and the laws of motion.

*Einstein field equations describe the gravity and the motion of matter objects, but they do not explain their physical nature.*

*The physical reality (Universe) is (must be) differentiated, i.e. in the final result it is (must be) created by the particles.*

If the physical reality were not differentiated (i.e. if it were not created by particles), it would be fused into an abstract *absolute identity* (*singularity*, i.e. into an abstract *mathematical point*), and so would actually physically not exist.

*However, the physical reality can not be differentiated unlimitedly to arbitrarily small parts.* Because if it were differentiated without limits (*ad absurdum*), it would represent the reverse extreme: an abstract *absolute no-identity* (i.e. an abstract *Euclidean empty space*), and thus again would not really physically exist.

*The physical reality (Universe) is (must be) differentiated, and simultaneously continuous.*

*The differentiated-continuous physical reality (Universe) is created by the particles* (more precisely *fluctuated particles*) *that interrelate to each other by interactions.*

The interaction of particles can be realized only in two ways [16]:

1. Mediated by the fluctuating exchange particles (gauge bosons): electromagnetic, weak nuclear and strong nuclear interactions.

2. Permanent fluctuations of all real and virtual particles in the local space-time (field): the inertial-gravitational relativistic interactions.

From these facts it results unambiguously: *All real and virtual particles have quantum-relativistic properties.*

*The result of the quantum-relativistic dynamics of the particles are relativistic (i.e. inertial-gravitational), properties of the micro-world and the macro-world, too.*

We can verify the fact that all micro-physical and macro-physical effects are the result of the quantum-relativistic dynamics of particles and have relativistic properties by:

A) Using present top observational technology.

B) An analysis of the results of sufficiently precise observations made in the past.

## 3 Conclusions

From the above analysis of the correlation of quantum and relativistic properties of particles it results unambiguously: The unified quantum-relativistic physical reality (Universe) is created by the permanent fluctuations of all real and virtual particles in the local space-time (field), as a result of the Heisenberg uncertainty principle.

The above analysis *de facto* confirms the prophetic words of Steven Weinberg, which we cited as a motto at the beginning of this article. Therefore, let us remind them again. Now, however, in the shorter and more radical form: "…the world is what we've always known: the Standard Model and general relativity." [17].